\documentclass[preprint,showpacs,preprintnumbers,amsmath,amssymb,superscriptaddress]{revtex4}
\usepackage{graphicx,bm}

\newcommand{\nl}{\nonumber \\}
\newcommand{\be}{\begin{equation}}
\newcommand{\ee}{\end{equation}}
\newcommand{\bea}{\begin{eqnarray}}
\newcommand{\eea}{\end{eqnarray}}
\newcommand{\bsube}{\begin{subequations}}
\newcommand{\esube}{\end{subequations}}
\newcommand{\Fig}[1]{Fig.\,\ref{#1}}
\newcommand{\Eq}[1]{Eq.\,(\ref{#1})}
\newcommand{\Eqs}[1]{Eqs.\,(\ref{#1})}
\newcommand{\alf}{\alpha}

\newcommand{\omg}{\omega}

\newcommand{\Gam}{\Gamma}
\newcommand{\gam}{\gamma}

\newcommand{\vpl}{\varepsilon}
\newcommand{\epl}{\epsilon}
\newcommand{\Dlt}{\Delta}

\newcommand{\la}{\langle}
\newcommand{\ra}{\rangle}
\newcommand{\ti}{\tilde}

\newcommand{\rmd}{{\rm d}}
\newcommand{\rmw}{{\rm w}}

\newcommand{\clT}{{\cal T}}
\newcommand{\Tr}{{\rm Tr}}
\newcommand{\Lp}{{\gamma_+}}
\newcommand{\Lm}{{\gamma_-}}
\newcommand{\bLp}{{\bar\gamma_+}}
\newcommand{\bLm}{{\bar\gamma_-}}

\begin{document}

\title{Reduced dynamics with renormalization
 in solid-state charge qubit measurement}

 \author{JunYan Luo}
 \email{firstluo@semi.ac.cn}
 \affiliation{Department of Chemistry,
 Hong Kong University of Science and
 Technology, Kowloon, Hong Kong}

 \author{Hujun Jiao}
 \affiliation{State Key Laboratory for
 Superlattices and Microstructures,
 Institute of Semiconductors,
 Chinese Academy of Sciences,
 P.O.~Box 912, Beijing 100083, China}

 \author{Feng Li}
 \affiliation{State Key Laboratory for
 Superlattices and Microstructures,
 Institute of Semiconductors,
 Chinese Academy of Sciences,
 P.O.~Box 912, Beijing 100083, China}

 \author{Xin-Qi Li}
 \affiliation{Department of Chemistry,
 Hong Kong University of Science and
 Technology, Kowloon, Hong Kong}
 \affiliation{State Key Laboratory for
 Superlattices and Microstructures,
 Institute of Semiconductors,
 Chinese Academy of Sciences,
 P.O.~Box 912, Beijing 100083, China}
 \affiliation{Department of Physics,
 Beijing Normal University,
 Beijing 100875, China}

 \author{YiJing Yan}
 \affiliation{Department of Chemistry,
 Hong Kong University of Science and
 Technology, Kowloon, Hong Kong}

\begin{abstract}
 Quantum measurement will inevitably cause backaction on
 the measured system, resulting in the well known dephasing and
 relaxation. In this report, in the context of solid--state
 qubit measurement by a mesoscopic detector, we show that
 an alternative backaction known as \emph{renormalization}
 is important under some circumstances. This effect
 is largely overlooked in the theory of quantum measurement.
\end{abstract}


\maketitle

\section{Introduction}

 One of the key requirements for physically implementing
 quantum computation is the ability to readout a
 two--state quantum system (qubit).
 Among various proposals, an important one is to
 use an electrometer as detector whose conductance
 depends on the charge state of a nearby qubit.
 Such electrometer can be a quantum point contact
 (QPC) \cite{Ale973740,Gur9715215,Kor01165310,%
 Goa01125326,Ave05126803,Pil02200401,Cle03165324,%
 Li05066803}, or a single
 electron transistor \cite{Shn9815400,Dev001039,%
 Mak01357,Cle02176804,Jia07155333,Gil06116806,%
 Gur05073303,Oxt06045328}.
 Both of them have been preliminarily implemented
 in experiment for quantum
 measurements \cite{Buk98871,Sch981238,Nak99786,%
 Spr005820,Aas013376}.
 Also, similar structures were proposed for
 entanglement generation and detection by conduction
 electrons \cite{Elz04431,Tra06235331,Sch0767}.

 The problem of measuring a charge qubit by a QPC
 detector has been well studied in high bias voltage
 regime.
 Work for arbitrary measurement voltage has also
 been reported although relatively
 limited \cite{Li05066803,Shn0211618}.
 Most of them only dealt with the measurement
 induced dephasing and relaxation, which, from
 the perspective of information, are consequences
 of information acquisition by measurement.
 The physical interaction between the measurement
 apparatus and the qubit, however, give rise to
 another important backaction which renormalizes the
 internal structure of qubit.

 In this context, we revisit the measurement problem,
 while take fully into account of the energy
 renormalization.
 This effect was often disregarded in the
 literature.
 Indeed, the steady--state renormalization can be
 effectively included in the Caldeira--Leggett
 renormalized system
 Hamiltonian \cite{Xu029196,Yan05187,Cal83587,Wei08}.
 The resulting dynamics is however different in detail
 from that of the dynamical
 renormalization approach \cite{Xu029196,Yan05187}. The
 apparent distinction should be sensitively reflected
 in the output power spectral density studied in this work.
 Our analysis shows that the renormalization effect
 on qubit becomes increasingly important as one lowers
 the measurement voltage. Therefore, it would
 require in practice to have this feature being taken
 into account properly, in order to correctly analyze and
 understand the measurement results.

 \begin{figure}
 \begin{center}
 \includegraphics*[scale=0.80]{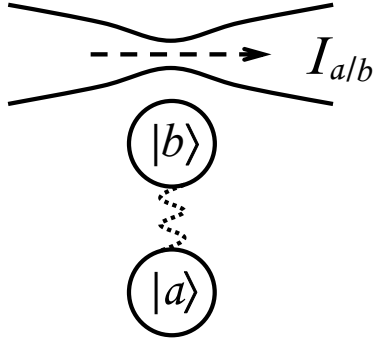}
 \caption{\label{fig1}
 Schematic setup of a solid--state charge qubit
 measured continuously by a quantum--point--contact (QPC).}
 \end{center}
 \end{figure}

 \section{Model description}

 The system under investigation is schematically shown in
 \Fig{fig1}. The Hamiltonian of the entire system is of
 $H_{\rm T} =H_{\rm qu}+H_{\rm D}+H'$, with the qubit,
 QPC detector, and their coupling parts being modeled
 respectively by
 \bsube\label{sys-Ham}
 \bea
 &&H_{\rm qu} = \sum_{s=a,b}\epl_s|s\ra\la s|
 +\case{1}{2}\Delta
 \left(|a\ra\la b|+|b\ra\la a|\right),\label{Hqubit}
 \\
 &&H_{\rm D} =\sum_{k\in {\rm L}} \vpl_k
 \hat c_k^\dag \hat c_k
 +\sum_{q\in {\rm R}} \vpl_q
 \hat c_q^\dag \hat c_q \, ,
 \\
 &&H' =\sum_{s=a,b}\sum_{k,q}t^{s}_{kq}
 \hat c_k^\dag \hat c_q
 \cdot |s\ra\la s|+{\rm H.c.}\label{Hprim}
 \eea
 \esube
 The amplitude $t^s_{kq}$ of electron tunneling
 through two reservoirs ($\alf=$ L and R) of the
 QPC depends explicitly on the qubit state.
 Denote $Q_s\equiv |s\ra\la s|$ hereafter.
 Thus, the qubit--QPC detector coupling reads
 in the $H_{\rm D}$--interaction picture as
 $H'(t)=\sum_s[\hat f_s(t)+\hat f^{\dag}_s(t)]\cdot Q_s$,
 with
 $\hat f_s(t)\equiv e^{iH_{\rm D}t}\big(\sum_{kq}t_{kq}^s
 \hat c_k^\dag \hat c_q\big)e^{-iH_{\rm D}t}$.
 The effects of the stochastic QPC reservoirs on
 measurement are characterized by
 $\ti C_{ss'}^{(+)}(t-\tau)\equiv\la \hat f_{s}^\dag(t)
 \hat f_{s'}(\tau)\ra$ and
 $\ti C_{ss'}^{(-)}(t-\tau)\equiv\la \hat f_{s}(t)
 \hat f_{s'}^\dag(\tau)\ra$.
 In terms of the reservoirs spectral density functions,
 which are defined physically as
 \be\label{Jww}
 J_{ss'}(\omg,\omg')=\sum_{k,q} t_{kq}^st_{kq}^{s'}
 \delta(\omg-\vpl_k)\delta(\omg'-\vpl_q),
 \ee
 these QPC coupling correlation functions are
\[
 \ti C_{ss'}^{(\pm)}(t) = \! \int\!\!\!\int\!\! d\omg d\omg'
 J_{ss'}(\omg,\omg')f_{\rm L}^{(\pm)}(\omg)
 f_{\rm R}^{(\mp)}(\omg') e^{i(\omg-\omg')t} .
\]
 Here,
 $f_\alf^{(\pm)}(\omg)=\{1+e^{\pm\beta(\omg-\mu_{\alf})}\}^{-1}$
 relates to the Fermi function of the lead $\alf$, with
 $\beta=(k_{\rm B}T)^{-1}$ the inverse temperature.
 The coupling spectrum function used later is defined by
 \be \label{Cssw}
 C_{ss'}^{(\pm)}(\omg)\equiv \int_{-\infty}^{\infty}
 dt \ti C_{ss'}^{(\pm)}(t)e^{-i\omg t}.
 \ee
 Throughout this work, we set
 $\mu^{\rm eq}_{\rm L}=\mu^{\rm eq}_{\rm R}=0$ for the
 equilibrium chemical potentials  (or Fermi energies)
 of the QPC reservoirs in the absence of applied bias
 voltage, and $\hbar = e =1$ for the Planck constant
 and electron charge.

\section{Particle-number-resolved master equation}

 The reduced density matrix of the qubit is formally
 defined as
 $\rho(t)\equiv {\rm Tr}_{\rm D}[\rho_{\rm T}(t)]$,
 i.e., tracing out the QPC reservoirs degree of
 freedom over the entire qubit--plus--detector
 density matrix.
 The qubit system Liouvillian is defined via
 ${\cal L}\hat O  \equiv [H_{\rm qu}, \hat O]\,$. By
 treating $H'$ as perturbation, a master equation for
 the reduced density matrix can be derived
 as \cite{Xu029196,Yan05187,Yan982721}
 \be\label{master0}
 \dot{\rho}(t)=-i{\cal L}\rho(t)-\frac{1}{2}\sum_s
 [Q_s,\tilde{Q}_s\rho(t)-\rho(t)\tilde{Q}_s^\dag],
 \ee
 with $\ti{Q}_s\equiv \ti{Q}_s^{(+)}+\ti{Q}_s^{(-)}$,
 and
 \be\label{tiQpm}
 \ti{Q}_s^{(\pm)}\equiv \sum_{s'}
 [{C}_{ss'}^{(\pm)}({\cal L})+i D_{ss'}^{(\pm)}({\cal L})]
 {Q_{s'}}.
 \ee
 Here, $C_{ss'}^{(\pm)}({\cal L})\equiv C_{ss'}^{(\pm)}
 (\omg)|_{\omg={\cal L}}$ is the spectrum function
 defined earlier. The dispersion function
 ${D}_{ss'}^{(\pm)}({\cal L})$ can then be evaluated via
 the Kramers-Kronig relation,
 \be \label{dispersion}
 D_{ss'}^{(\pm)}(\omg)=\frac{1}{\pi}\,{\cal P}
 \!\!\int_{-\infty}^{\infty} d\omg'
 \frac{C_{ss'}^{(\pm)}(\omg')}{\omg-\omg'} .
 \ee
 Physically, it is responsible for the
 renormalization \cite{Xu029196,Yan05187,Cal83587,Wei08}.

 To achieve a description of the output from detector, we
 employ the transport particle number ``$n$''-resolved
 reduced density matrices \{$\rho^{(n)}(t); n=0,1,\cdots$\}
 that satisfy $\rho(t)=\sum_n \rho^{(n)}(t)$. The
 corresponding ``$n$''-resolved conditional quantum master
 equation reads \cite{Li05066803,Li05205304,Luo07085325}
 \bea\label{CQME}
 \dot{\rho}^{(n)}(t)&= -i{\cal L}\rho^{(n)}(t)
 -\frac{1}{2}\sum_{s}\big\{Q_s\ti{Q}_s\rho^{(n)}
 \! -\ti{Q}^{(-)}_s\rho^{(n-1)}Q_s
 \nl
 &\quad\quad
 -\ti{Q}^{(+)}_s\rho^{(n+1)}Q_s+{\rm H.c.}\big\}.
 \eea

 We would like to account for the finite bandwidth of the
 QPC detector, which will be characterized by a single
 Lorentzian.
 Real spectral density has a complicated structure,
 which can be parameterized via the technique of spectral
 decomposition \cite{Mei993365,Li07075114}.
 This complexity, however, will only modify details of the
 results, but not the qualitative picture.
 For the sake of constructing analytical results, we assume
 a simple Lorentzian function centered at the Fermi energy
 for the spectral density \Eq{Jww}.
 This choice stems also from the assumption that the energy
 band of each reservoir is half--filled.
 Moreover, the bias voltage is conventionally described by
 a relative shift of the entire energy-bands, thus the
 centers of the Lorentzian functions would fix at the
 Fermi levels. Without loss of generality, we simply
 assume
 \be\label{Jw}
 J_{ss'}(\omg,\omg')=\clT_s\clT_{s'}
 \frac{\Gam_{\rm L}^0\rmw^2}{(\omg-\mu_{\rm L})^2+\rmw^2}\cdot
 \frac{\Gam_{\rm R}^0\rmw^2}{(\omg'-\mu_{\rm R})^2+\rmw^2}.
 \ee
 We set $\clT_a \equiv 1$ and $\clT_b \equiv 1 -\chi$.
 The asymmetric qubit--QPC coupling parameter is of
 $0<\chi < 1$, as inferred from \Fig{fig1}.
 The correlation function of \Eq{Cssw} can be evaluated
 as $C_{ss'}^{(\pm)}(\omg)=\clT_s\clT_{s'}
 C^{(\pm)}(\omg)$, with
 \bea\label{Cpm}
 C^{(\pm)}(\omg)\!=\frac{\eta g(x)}
 {1-e^{\beta x}}
 \!\!\left[\!\frac{\rmw^2}{x}
 \{\phi(0)\!-\!\phi(x)\}\!-\!\frac{\rmw}{2}\varphi(x)
 \right]_{x=\omg\pm V}.
 \eea
 Here, $\eta=2\pi\Gam_{\rm L}^0\Gam_{\rm R}^0$,
 $g(x)=4\rmw^2/(x^2+4\rmw^2)$, and
 $V=\mu_{\rm L}-\mu_{\rm R}$ the applied voltage on the
 QPC detector;
 $\phi(x)$ and $\varphi(x)$ denote the real and imaginary
 parts of the digamma function
 $\Psi(\frac{1}{2}+\beta\frac{\rmw+ix}{2\pi})$, respectively.
 Knowing the spectral function, the dispersion function
 $D_{ss'}^{(\pm)}(\omg)=\clT_s \clT_{s'}D^{(\pm)}(\omg)$
 can be obtained via the Kramers--Kronig relation.
 The present spectrum functions satisfy
 the detailed--balance relation
 $C^{(+)}(\omg)=e^{-\beta(\omg+V)}C^{(-)}(-\omg)$.
 This means that our approach properly accounts for
 the energy exchange between the qubit and the detector
 during  measurement.

\section{Output power spectral density}

 In continuous weak measurement of qubit oscillations,
 the most important output is the spectral density of
 current. Typically, the power spectrum is defined with
 a stationary state. The involving stationary--state
 $\rho^{\rm st}$ can be determined by setting
 $\dot\rho^{\rm st}=0$ in \Eq{master0}, together with
 the normalization condition, at given bias voltage and
 temperature.
 For clarity, we focus hereafter on the symmetric
 qubit case, with the state energies of
 $\epl_a=\epl_b = 0$.

 Let us start with the average current. Using the
 ``$n$''-resolved master equation (\ref{CQME}), the
 average current can be expressed as
 $I(t)=\sum_n n {\rm Tr}[\dot{\rho}^{(n)}(t)]
 ={\rm Tr}[{\cal J}^{(-)}\rho(t)] $,
 where ${\cal J}^{(-)}$ is one of the superoperators,
 defined as
 \bea
 {\cal J}^{(\pm)}\rho(t)\equiv\frac{1}{2}
 \sum_s\big(\ti{Q}_s^{(-)}\pm\ti{Q}_s^{(+)}\big)
 \rho(t)Q_s+{\rm H.c.}
 \eea
 The stationary current can be carried out as
 \be\label{barI}
 \bar{I}=I_a\rho_{aa}^{\rm st}+I_b\rho_{bb}^{\rm st}
 +I_{ab}\rho_{ab}^{\rm st},
 \ee
 which for a symmetric qubit ($\epl_a=\epl_b =0$) is of
 \bea\label{allI}
 I_a &=&(1-\case{\chi}{2})
 C(0)\tanh(\case{\beta V}{2}) +\chi\bLp  \, ,
\nl
 I_b&=&(1-\chi)\left[(1-\case{\chi}{2})
 C(0)\tanh(\case{\beta V}{2}) -\chi\bLp\right]  \, ,
\\
 I_{ab}&=&\chi^2\bLm \,  .
 \nonumber
 \eea
 Here, $\bar\gamma_{\pm}\equiv\case{1}{4}\big[\bar C(\Dlt)
 \pm\bar C(-\Dlt)\big]$, with
 $\bar C(\omg)\equiv C^{(-)}(\omg)-C^{(+)}(\omg)$. Denote
 also $C(\omg)\equiv C^{(-)}(\omg)+C^{(+)}(\omg)$.

 The noise spectral density can be calculated via the
 MacDonald's formula \cite{Mac62}
 \be \label{MacD}
 S(\omg)=2\omg\int_0^\infty\!\!dt\sin(\omg t)
 \frac{d}{dt}[\la n^2(t)\ra-(\bar{I}t)^2],
 \ee
 with $\la n^2(t)\ra\equiv\sum_nn^2\Tr\{\rho^{(n)}(t)\}$.
 Applying equation (\ref{CQME}) gives
 \be\label{n2_eom}
 \frac{d}{dt}\la n^2(t)\ra={\rm Tr}
 [2{\cal J}^{(-)}N(t)+{\cal J}^{(+)}\rho^{\rm st}],
 \ee
 where $N(t)\equiv\sum_n n\rho^{(n)}(t)$, which can
 be calculated via
 \be\label{EOM-N}
 \frac{dN}{dt}=-i\mathcal{L}N-\frac{1}{2}\sum_s
 \big[Q_s,\ti{Q}_s N-N\ti{Q}_s^\dag\big]+
 {\cal J}^{(-)}\rho(t).
 \ee
 For a symmetric qubit, analytical result is available.
 We split the spectrum into four components,
 $ S=S_0+S_1+S_2+S_3 $,
 and present them one--by--one as follows. First, the
 frequency-independent background noise $S_0$ reads
 \bea\label{S0}
 S_0 &=& 2\bar{I}\coth(\textstyle\frac{\beta V}{2})
 -\chi^2(\Lm/\Lp)
 \big[\Lm \!-\! \bLm
 \coth(\textstyle\frac{\beta V}{2})\big]
 \nl
 &&\quad
 +\chi\big[\chi -(2-\chi)\delta{\bar P}\big]
 \big[\Lp \!-\! \bLp
 \coth(\textstyle\frac{\beta V}{2})\big],
 \eea
 with $\gamma_{\pm}\equiv\case{1}{4}\big[C(\Dlt)
 \pm C(-\Dlt)\big]$ and $\delta{\bar P}\equiv
 \rho^{\rm st}_{bb}-\rho^{\rm st}_{aa}$ that is
 nonzero due to the asymmetric qubit--QPC coupling.
 The second component is a Lorentzian, with the
 peak at $\omg=0$ and the dephasing rate of
 $\gam_\rmd=\chi^2\Lp$. It reads
 \be\label{S1}
 S_1=(X\gam_\rmd-\chi^2\Lm\bar{I})
 \frac{2I_{ab}}{\omg^2+\gam_\rmd^2}.
 \ee
 Here,
 $2\la a|{\cal J}^{(+)}\rho^{\rm st}|b\ra\equiv X+iY$
 (the real and imaginary parts).
 We remark that $S_1$ arises completely from the qubit
 relaxation induced inelastic tunneling effect in the
 detector\cite{Li05066803}.
 The last two components are
 \bea
 S_2 &=&\left[\!\frac{(\chi^2\Lm
 \bar{I}\!-\!X\gam_\rmd)\tilde{\epsilon}}
 {\omg^2+\gam^2_\rmd}
\! + \! Y\right]
 \frac{\omg^2 {\gam}'_\rmd A\!-\!(\omg{\omg'}
 \!-\!\Dlt\tilde{\Delta}) B}{\omg^2{\gam'_\rmd}^2
 +(\omg{\omg'}\!-\!\Dlt\tilde{\Delta})^2},
 \label{S2}
\\
 S_3 &=& \left[\!\frac{(\chi^2\Lm
 \bar{I}\gam_\rmd\!+\!X\omg^2)\tilde{\epsilon}}
 {\omg^2+\gam^2_\rmd} \!-\! \Dlt Z\right]
 \frac{\gam'_\rmd B\!-\!(\omg{\omg'}
 \!-\!\Dlt\tilde{\Delta}) A}{\omg^2{\gam'_\rmd}^2
 \!+\!(\omg{\omg'}\!-\!\Dlt\tilde{\Delta})^2}\, .\label{S3}
 \eea 
 Here, $\tilde{\epsilon}$ and $\tilde{\Delta}$ are the
 renormalized version of the original
 $\epl\equiv\epl_a-\epl_b$ and $\Delta$ of the qubit.
 They are related to the dispersion functions of the
 detector. Let
 $D(\omg)\equiv D^{(-)}(\omg)+D^{(+)}(\omg)$ and
 $\bar D(\omg)\equiv D^{(-)}(\omg)-D^{(+)}(\omg)$.
 Simple analysis on the symmetric case ($\epsilon=0$)
 gives
 \bsube\label{tieps_Del}
 \bea
 {\ti\epsilon}&=&\chi(1-\case{\chi}{2}) D(0)\, ,
 \\
 \ti{\Delta}&=&\Delta+\case{1}{4}\chi^2[D(\Dlt)-D(-\Dlt)]\, .
 \eea
 \esube
 For the bookkeeping of \Eqs{S2} and (\ref{S3}), we
 have also introduced
 $Z\equiv\textstyle{\frac{1}{2}}[(I_a+I_b)\delta{\bar P}
 +(I_a-I_b)]+(\case{2}{\chi}-1)I_{ab}\rho_{ab}^{\rm st}$,
 and the frequency--dependent quantities of
 \bea\label{para_on_w}
 &&\omg' \equiv
 \omg\big(1-\textstyle{\frac{\tilde{\epsilon}^2}
 {\omg^2+\gam^2_\rmd}}\big)\, ,
 \qquad\;
 \gam'_\rmd \equiv
 \gam_\rmd\big(1+\textstyle{\frac{\tilde{\epsilon}^2}
 {\omg^2+\gam^2_\rmd}}\big)\, ,
 \nl
 &&A \equiv \chi(1-\case{\chi}{2})
 [\bar D(\Dlt) - \bar D(-\Dlt)]
 +2\tilde{\epsilon}I_{ab}
 \case{\gam^2_\rmd}{\omg^2+\gam^2_\rmd}\, ,
\\
 &&B \equiv -2(I_a-I_b)\Dlt+2\tilde{\epsilon}I_{ab}
 \case{\omg^2}{\omg^2+\gam^2_\rmd}\, .
 \nonumber
 \eea

 \begin{figure}
 \begin{center}
 \includegraphics*[scale=0.70]{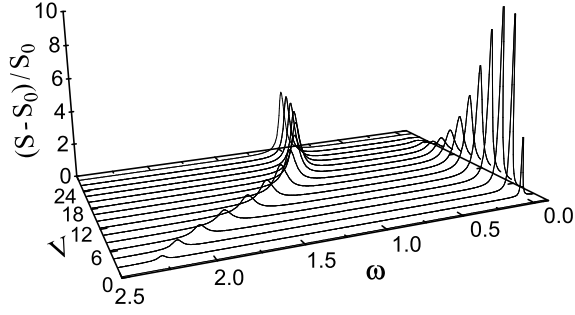}
 \caption{\label{fig2}
 Power spectral density of the detector current,
 with the frequency and voltage labeled in unit of $\Delta$.
 The bandwidth $\rmw=15\Dlt$.
 Other parameters are
 $\eta=2$, $\chi=0.2$ and $\beta\Dlt=1$.}
 \end{center}
 \end{figure}

 The computed noise spectrum is displayed in \Fig{fig2}.
 It is of interest to note that the spectral peak apart
 from the zero frequency, which is the signal of qubit
 oscillations, shifts with the measurement voltages.
 (i) In the high voltage regime  (e.g.\ for
 $V\gtrsim 30\,\Dlt$ as shown in \Fig{fig2}), the
 oscillation peak locates approximately at
 $\omg\approx \Dlt$;
 (ii) As lowering the voltage, the measurement induced
 renormalization effect becomes increasingly important,
 which strongly affects the position of the oscillation
 peak.

\begin{figure}
 \begin{center}
 \includegraphics*[scale=0.56]{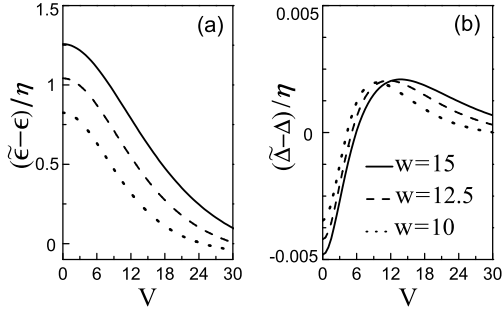}
 \caption{\label{fig3}
 Renormalization on the qubit level energy (a) and
 coupling (b), exemplified with three values of
 bandwidth $\rmw$ (in unit of $\Delta$). Other
 parameters are the same as in \Fig{fig2}. }
 \end{center}
 \end{figure}

 The feature of the noise spectrum in \Fig{fig2} is
 closely related to the renormalization of the qubit
 parameters $\epl$ and $\Delta$. In the limit of weak
 qubit-QPC coupling, the renormalized Rabi frequency
 is given by
 $\omg_{\rm R}=\sqrt{{\tilde{\epsilon}}^2+\ti\Delta^2}$.
 The renormalization
 effect ($\omg_{\rm R}-\sqrt{\epsilon^2+\Delta^2}$)
 increases {\it monotonically} with the QPC bandwidth
 (w). In \Fig{fig3} we plot $\tilde{\epsilon}$ and
 $\ti{\Delta}$, in terms of the $\eta$--scaled
 renormalizations, against the bias voltage for
 different bandwidths.
 The renormalized qubit state energy difference
 $\tilde{\epsilon}$ increasingly deviates from the
 original $\epsilon=0$ as the QPC bandwidth increases
 or the applied voltage decreases, as shown in
 \Fig{fig3}(a).
 In contrast, the inter-state coupling renormalization
 is negligibly small, as depicted in \Fig{fig3}(b)
 and also claimed in Ref.\ \cite{Shn0211618}.
 That $(\tilde{\epsilon}-\epl)$ being dominant can
 be readily understood by the form of coupling $H'$
 of \Eq{Hprim}, which modulates the level energies,
 rather than the level coupling.
 In the wideband limit ($\rmw\rightarrow\infty$),
 the energy renormalization would diverge.
 However, this feature is an artifact, since in
 reality a natural cutoff of the bandwidth must exist.
 That's the reason we introduce a Lorentzian cut--off
 in \Eq{Jw}.

 The noise spectrum itself depends on
 $\eta$ in a rather complicated manner, especially
 the $S_2$ and $S_3$ components [\Eq{S2} or
 \Eq{S3} with \Eq{para_on_w}] that are of
 dynamical in nature.
 In contrast, the algebraic nonlinear dependence
 of $\eta$ in the average current $\bar I$
 \Eq{barI} and $S_0$ \Eq{S0} arises from the
 renormalized stationary $\rho^{\rm st}$ only.
 In literature (e.g.\ Ref.\ \cite{Shn0211618}) the
 dispersion function is often disregarded
 explicitly, with its effect being included in
 the Caldeira-Leggett renormalized system
 Hamiltonian\cite{Xu029196,Yan05187,Cal83587,Wei08}.
 However, this approach gives rise to quite
 different dynamics from the present result,
 even though their stationary state behaviors
 could be similar \cite{Xu029196,Yan05187}.
 Apparently, the dynamical distinct should be
 sensitively reflected in the shot noise spectrum.
 In the context of qubit measurement by a QPC
 detector, our analysis can be served as a
 detailed investigation of the dynamical
 renormalization effect.

\begin{figure}
 \begin{center}
 \includegraphics*[scale=0.60]{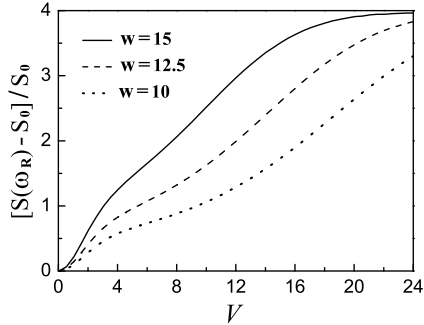}
 \caption{\label{fig4}
 Bias voltage dependence of the peak-to-pedestal ratio of
 the output power spectrum, exemplified with the same
 parameters in \Fig{fig3}.}
 \end{center}
 \end{figure}

 In \Fig{fig4} we further show the
 signal--to--noise ratio of the noise spectrum
 against the bias voltage for different bandwidths.
 In the limit of large bias $V\gg\Dlt$ and for weak
 qubit-QPC coupling, the signal--to--noise ratio
 \be\label{SNlimit}
 \frac{S(\omg_{\rm R})-S_0}{S_0}\bigg|_{V\gg\Dlt}
 \rightarrow 4\, \frac{(2-\chi)^2}
 {(2-\chi)^2+\chi^2}
 \ee
 can reach the limit of 4; i.e.\ the Korotkov--Averin
 bound for any linear response
 detectors \cite{Kor01165310,Kor01085312,Ave05069701,Jia08062502}.

 As seen in \Fig{fig2}, the detector induced
 renormalization also results in a wide voltage
 range where the coherent peak at the
 renormalized Rabi frequency and the sharp peak
 at zero frequency coexist.
 In that regime, the level mismatch induced by
 the detector is prominent, while the qubit
 coherence is not strongly destroyed.
 As is well known \cite{Kor01165310,%
 Shn0211618,Gur03066801}, the peak at zero frequency is a
 signature of the Zeno effect in continuous
 weak measurement. The basic picture is that
 the detector attempts to localize the electron
 in one of the levels for a longer time, leading
 thus to incoherent jumps between the two levels.
 Finally, in \Fig{fig2}, the coherent peak
 persists to high bias voltage, while the
 zero-frequency peak eventually disappears.
 This feature is different from the previous
 work \cite{Li05066803}.
 The reason is twofold. On one hand, as shown
 in \Fig{fig3}, the renormalization of energy
 levels is weak at high voltage. On the other
 hand, in this work we adopted a finite bandwidth
 model for the QPC. This implies that in high
 voltage regime the QPC (measurement) current is
 weak, which differs from the result under the
 usual wide-band approximation. As a consequence,
 the weak backaction from the detector together
 with the alignment of the qubit levels results
 in the spectral feature shown in \Fig{fig2} at
 high voltages.

 \section{Conclusions}

 In summary, we have revisited the problem of
 continuous measurement of a solid--state qubit
 by quantum point contact. Our results showed
 that the renormalization effect, which was
 neglected in previous studies, can
 significantly affect the output power spectrum.
 This feature should be taken into account in the
 interpretation of measurement result.
 We also note that the renormalization in the
 present setup may be quantified {\it in situ}.
 No reference to the bare qubit is needed, as
 it can be effectively replaced the band--edge
 large voltage transport limit.

\begin{acknowledgments}
 This work is supported by RGC (604007 \& 604508) of
 Hong Kong SAR Government, the National Natural Science
 Foundation (60425412 \& 90503013), and Major State
 Basic Research Project (2006CB921201) of China.
\end{acknowledgments}


\end{document}